\documentclass[11pt,amsfonts]{article}
\usepackage{amssymb}

\newcommand{\neqq}{\equiv\!\!\!\!\!\!/\ }

\newcommand{\ad}{\mbox{$ad$ }}

\newcommand{\be}{\begin{equation}}
\newcommand{\ee}{\end{equation}}
\newcommand{\bea}{\begin{array}}
\newcommand{\ea}{\end{array}}
\newcommand{\beqa}{\begin{eqnarray}}
\newcommand{\eeqa}{\end{eqnarray}}
\newcommand{\bean}{\begin{eqnarray*}}
\newcommand{\eean}{\end{eqnarray*}}

\def\up#1{\leavevmode \raise.16ex\hbox{#1}}

\newcommand{\gapproxeq}{\lower
 .7ex\hbox{$\;\stackrel{\textstyle >}{\sim}\;$}}
\newcommand{\lapproxeq}{\lower .7ex\hbox{$\;\stackrel
{\textstyle <}{\sim}\;$}}

\newcounter{appendice}

\def\thebibliography#1{{\bf REFERENCES\markboth
 {REFERENCES}{REFERENCES}}\list
 {[\arabic{enumi}]}{\settowidth\labelwidth{[#1]}\leftmargin\labelwidth
 \advance\leftmargin\labelsep
 \usecounter{enumi}}
 \def\newblock{\hskip .11em plus .33em minus -.07em}
 \sloppy
 \sfcode`\.=1000\relax}

\begin{document}
\begin{flushright}
 SU-4252-802\\
 03-20-2005
\end{flushright}

\vskip 1cm

\centerline{ \LARGE On Time-Space Noncommutativity for Transition
Processes} \centerline{ \LARGE and} \centerline{ \LARGE
Noncommutative Symmetries}

\vskip 2cm

\centerline{ {\sc    A.P. Balachandran\footnote{e-mail:
bal@physics.syr.edu} and A. Pinzul\footnote{e-mail:
apinzul@physics.syr.edu} } }

\vskip 1cm
\begin{center}
{\it  Department of Physics, Syracuse University,\\ Syracuse, New
York 13244-1130,  USA}

\end{center}

\vskip 2cm

\vspace*{5mm}

\normalsize \centerline{\bf ABSTRACT}

We explore the consequences of time-space noncommutativity in the
quantum mechanics of atoms and molecules, focusing on the Moyal
plane with just time-space noncommutativity ($[\hat{x}_\mu ,
\hat{x}_\nu]=i\theta_{\mu\nu}$, $\theta_{0i}\neqq 0$,
$\theta_{ij}=0$). Space rotations and parity are not automorphisms
of this algebra and are not symmetries of quantum physics. Still,
when there are spectral degeneracies of a time-independent
Hamiltonian on a commutative space-time which are due to
symmetries, they persist when $\theta_{0i}\neqq 0$; they do not
depend at all on $\theta_{0i}$. They give no clue about rotation
and parity violation when $\theta_{0i}\neqq 0$. The persistence of
degeneracies for $\theta_{0i}\neqq 0$ can be understood in terms
of invariance under deformed noncommutative ``rotations'' and
``parity''. They are not spatial rotations and reflection. We
explain such deformed symmetries. We emphasize the significance of
time-dependent perturbations (for example, due to time-dependent
electromagnetic fields) to observe noncommutativity. The formalism
for treating transition processes is illustrated by the example of
nonrelativistic hydrogen atom interacting with quantized
electromagnetic field. In the tree approximation, the
$2s\rightarrow 1s +\gamma$ transition for hydrogen is zero in the
commutative case. As an example, we show that it is zero in the
same approximation for $\theta_{0i}\ne 0$. The importance of the
deformed rotational symmetry is commented upon further using the
decay $Z^0 \rightarrow 2\gamma$ as an example.

\vspace*{5mm}

\newpage
\scrollmode

\section{Introduction}

Different approaches to fundamental physics suggest space-time
noncommutativity. It arises naturally in quantum gravity when one
considers the localization of points in space-time
\cite{Doplicher:1994tu}. It arises in string theory as well in a
certain limit \cite{Seiberg:1999vs}.

Formulation of quantum physics on noncommutative space-times does
not present conceptual problems if time commutes with spatial
coordinates. That is not the case with time-space
noncommutativity. It was the work of Doplicher et al
\cite{Doplicher:1994tu} which systematically developed unitary
quantum field theories (QFT's) with time-space noncommutativity.
Their ideas were later adapted to quantum mechanics by
Balachandran et al \cite{Balachandran:2004rq}. It appears that we
now have the tools for doing consistent quantum physics with
time-space noncommutativity.

An important task is the extraction of observable consequences of
noncommutative space-times. An extensive literature already exists
on this subject for space-space noncommutativity (see
\cite{Hinchliffe:2002km} for a review and references), but that is
not the case for time-space noncommutativity. In this paper, we
make a beginning in this regard.

After reviewing previous work on quantum physics with time-space
noncommutativity, we consider certain implications of the spectral
map theorem of that work. The theorem states that if the
Hamiltonian has no explicit time-dependence, its spectra for
commutative and noncommutative space-times are identical, provided
only that spatial coordinates commute for the latter. For Moyal
space-times $\mathcal{A}_\theta (\mathbb{R}^{d+1})$, where
$$
[\hat{x}_\mu , \hat{x}_\nu]=i\theta_{\mu\nu}1\!\mbox{l}\ ,\ \
\theta_{\mu\nu} \mbox{ are real constants,}
$$
that means that \be \label{SpatCom} \theta_{ij}=0\ ,\ \
i,j\in[1,d] \ee  if $i,j$ denote the spatial and $0$ the time
operators. Now even with (\ref{SpatCom}), spatial rotations are
not automorphisms of $\mathcal{A}_\theta (\mathbb{R}^{d+1})$ if
$\theta_{0i}\neqq 0$. Nor is parity\footnote{We use the term
``parity'' as the reflection $\hat{x}_i\rightarrow -\hat{x}_i$ of
\textit{all} spatial coordinates when $d$ is odd. This is the
conventional definition. But we can also define it as a fixed
element of the orthogonal group $O(d)$ not connected to identity,
such as the reflection of a coordinate perpendicular to
$\vec{\theta}_0=(\theta_{01},...,\theta_{0d})$. We note that this
particular reflection \textit{is} an automorphism of our algebra.}
an automorphism if $d$ is odd.\footnote{For $d$ even, parity is
total reflection in the plane perpendicular to $\vec{\theta}_0$.}
Nevertheless for time-independent Hamiltonians invariant under
rotations or parity for $\theta_{\mu\nu}=0$, the spectral theorem
implies as a corollary that their energy degeneracies due to
symmetries remain intact when $\theta_{0i}\neqq 0$. Energy spectra
thus give no clue on noncommutative symmetry breakdown if the
Hamiltonian is time-independent. This is a surprising result. We
explain its conceptual reasons and emphasize the importance of
time-dependent phenomena for observing time-space
noncommutativity.

Next we develop the formulism for calculating transition processes
using the example of hydrogen atom interacting with
electromagnetic field. As an application, we consider the
$2s\rightarrow 1s + \gamma$ transition in hydrogen. It is
forbidden by parity if $\theta_{\mu\nu}=0$, but being a
time-dependent process, can occur if $\theta_{\mu\nu}\ne 0$. We
explicitly show that it is nevertheless zero in the tree
approximation.

The paper concludes with comments on possible effects of
time-space noncommutativity for processes such as $Z^0 \rightarrow
2\gamma$ which vanish by Poincar\'{e} invariance and Bose symmetry
in the commutative case \cite{Yang:1950rg},\cite{Behr:2002wx}.

\section{Quantum Mechanics on Noncommutative Space-Time}

This section is a short review of earlier work on the subject
\cite{Balachandran:2004rq}.

In the model we consider, $d=3$ and spatial coordinates commute,
\be \label{SpatCom1} [\hat{x}_i,\hat{x}_j]=0\ ,\ \ i,j\in[1,2,3]
\ee while there is time-space noncommutativity: \be
\label{CR0}[\hat{x}_0,\hat{x}_i]=i\theta s_i\ .\ee Here $s_i$ is a
fixed unit vector. We conveniently orient it in the 1-direction.
We thus assume that \be \label{CR} [\hat{x}_0,\hat{x}_1]=i\theta,\
\ [\hat{x}_0,\hat{x}_{2,3}]=0\ . \ee As $\theta\rightarrow
-\theta$ when $\hat{x}_0\rightarrow -\hat{x}_0$ (or
$\hat{x}_1\rightarrow -\hat{x}_1$), we further assume without loss
of generality that $\theta \geq 0.$

The algebra generated by $\hat{x}_\mu$ with the relations
(\ref{SpatCom1}) and (\ref{CR}) will be denoted by
$\mathcal{A}_\theta(\mathbb{R}^4)$.

In the algebraic approach to quantum physics, the quantum
mechanical Hilbert space $\mathcal{H}$ is built from elements of
$\mathcal{A}_\theta(\mathbb{R}^4)$ itself. Observables are
self-adjoint operators on this Hilbert space. In cases of interest
here, their domain contains
$\mathcal{A}_\theta(\mathbb{R}^4)\cap\mathcal{H}$.

Now the algebra itself acts in two distinct ways on
$\mathcal{A}_\theta(\mathbb{R}^4)$, namely by the left- and
right-regular representations
$\mathcal{A}_\theta^{L,R}(\mathbb{R}^4)$. For each $\hat{a}\in
\mathcal{A}_\theta(\mathbb{R}^4)$, we have $\hat{a}^{L,R}\in
\mathcal{A}_\theta^{L,R}(\mathbb{R}^4)$ where
$$
\hat{a}^L\hat\alpha=\hat{a}\hat{\alpha}\ ,\
\hat{a}^R\hat{\alpha}=\hat{\alpha} \hat{a}\ ,\
\hat\alpha\in\mathcal{A}_\theta(\mathbb{R}^4)\ .
$$
We can also associate the adjoint action $\ad\hat{a}$ to
$\hat{a}$:
$$
\ad{\hat{a}} = \hat{a}^L - \hat{a}^R\ ,\ \ad{\hat{a}}\,\hat\alpha
= [\hat{a},\hat\alpha]\ .
$$
Many observables of physical interest are obtained from
$\mathcal{A}_\theta^{L,R}(\mathbb{R}^4)$. In particular the
momentum $\hat{P}_1$ in 1-direction and the generator $\hat{P}_0$
of time translations are given by
$$
\hat{P}_1=-\frac{1}{\theta}\ad{\hat{x}^0}\ ,\
\hat{P}_0=-\frac{1}{\theta}\ad{\hat{x}^1}\ .
$$

As a preparation to construct the quantum Hilbert space, we next
introduce an inner product on $\mathcal{A}_\theta(\mathbb{R}^4)$.

Consider \be \label{Weyl1} \hat\alpha = \frac{1}{(2\pi)^2}\int d^4
p\, \tilde{a}(p)\mathrm{e}^{ip_i \hat x_i}\mathrm{e}^{ip_0 \hat
x_0}\ . \ee Its symbol $\alpha$ is a function
$\mathbb{R}^4\rightarrow \mathbb{C}$. We define it by \be
\label{Weyl2} \alpha(x) = \frac{1}{(2\pi)^2}\int d^4 p\,
\tilde{a}(p)\mathrm{e}^{ip_i x_i+ip_0 x_0}\ .\ee For the Moyal
symbol $\alpha_M$ of $\hat\alpha$, we would have written
$\mathrm{e}^{ip_i \hat x_i+ip_0 \hat x_0}$ in the RHS of
Eq.(\ref{Weyl1}), then $\alpha_M$ is the RHS of Eq.(\ref{Weyl2}).
Thus our $\alpha\ne \alpha_M$.

Using the symbol, we can define the positive map $S\ :\ \hat\alpha
\rightarrow\mathbb{C}$ \cite{Doplicher:1994tu} by
$$
S(\hat\alpha)=\int d^3 x\,\alpha (\vec{x},x_0)\ .
$$
The importance of $S$ is that it helps us to introduce an inner
product $(\cdot,\cdot)$ on $\mathcal{A}_\theta(\mathbb{R}^4)$: \be
\label{Scalar}(\hat\alpha , \hat\beta)= S(\hat\alpha^*
\hat\beta)=\int d^3 x\,\overline{\alpha (\vec{x},x_0)}\beta
(\vec{x},x_0)\ . \ee

The physical Hilbert space $\mathcal{H}$ is the (completion of
the) subspace of $\mathcal{A}_\theta(\mathbb{R}^4)$ subject to the
Schr\"{o}dinger equation (or constraint). Thus let $\hat H$ be a
Hamiltonian, Hermitian in the above inner product. Then if
${\hat\psi}\in\mathcal{A}_\theta(\mathbb{R}^4)\cap \mathcal{H}$,
\be \label{Constr} (\hat{P}_0-\hat{H})\hat\psi = 0\ . \ee

One can show that for vectors of $\mathcal{H}$, the above inner
product has no nontrivial null vectors and is also independent of
$x_0$.

The Hamiltonian is time-independent if
$$
[\hat{P}_0 , \hat{H}]=0\ .
$$
In that case the general solution of the Schr\"{o}dinger
constraint is \be \label{Solution} \hat\psi=\mathrm{e}^{-i\hat{H}
\hat{x}_0^R}\hat\varphi (\vec{\hat{x}})\ . \ee Here 1)
$\hat\varphi$ is time independent, $[\hat{P}_0 , \hat\varphi
(\vec{\hat{x}})]=0$, and 2) square-integrable,
$(\hat\varphi,\hat\varphi)<\infty$. We regard it as an element of
$\mathcal{A}_\theta(\mathbb{R}^4)$. Then $\hat{H}$ and
$\hat{x}_0^R$ act on it in Eq.(\ref{Solution}).

We can easily check that $\hat\psi$ fulfills (\ref{Constr}). Let
$H$ be a time-independent Hamiltonian in conventional quantum
physics with $\theta_{0i}=0$. It can be
$$
H = \frac{\vec{p}^2}{2m} + V(\vec{x})\ .
$$
Let $\varphi_E$'s be its eigenstates regarded as functions of
$\vec{x}$:
$$
H\varphi_E(\vec{x}) = E\varphi_E(\vec{x})\ .
$$
We can associate the Hamiltonian $\hat{H} =
H(\vec{\hat{P}},\vec{\hat{x}})$ to $H$ for $\theta_{0i}\ne 0$.
Then according to the spectral theorem, $\hat{H}$ and $H$ have
identical spectra while the eigenvectors of $\hat{H}$ are
${\hat\psi_E={\hat\varphi}_E}(\vec{\hat{x}})\exp (-iE \hat{x}_0)$:
$$
\hat{H}\hat\psi_E = E\hat\psi_E\ ,\ (\hat{P}_0 -
\hat{H})\hat\psi_E = 0\ .
$$
Proof is by inspection. It is important that $\hat\psi_E$ fulfills
the Schr\"{o}dinger constraint.

We refer to \cite{Balachandran:2004rq} for discussion of
time-dependent Hamiltonians.

\section{On Symmetries}

\textit{i) Commutative Rotations}

In commutative quantum physics where $\theta_{0i}= 0$, spatial
rotations are generated by angular momentum operators $L_i$ where
$$
L_i = \varepsilon_{ijk}x_j p_k\ ,\ p_k =
-i\frac{\partial}{\partial x_k}\ ,
$$
where $\varepsilon_{ijk}$ is the Levi-Civita symbol with
$\varepsilon_{123}=+1$. Spatial coordinates rotate under the
$SO(3)$ group generated by $L_i$, whereas time is a rotational
scalar:
\begin{eqnarray}
& &[L_i , x_j] = i \varepsilon_{ijk}x_k\ , \nonumber \\
& &[L_i , x_0] = 0\ . \nonumber
\end{eqnarray}
Momenta too rotate like $\vec{x}$:
$$
[L_i , p_j] = i \varepsilon_{ijk}p_k\ .
$$
These equations let us identify the $SO(3)$ group generated by
$L_i$ with spatial rotations.

\noindent\textit{ii) Noncommutative Rotations\footnote{For a
different approach to noncommutative space-time symmetries, see
\cite{Chaichian:2004za}.}}

For the algebra $\mathcal{A}_\theta(\mathbb{R}^4)$ as well, there
exist operators $\hat{L}_i$ which generate $SO(3)$:
\begin{eqnarray}
& &\hat{L}_i = \varepsilon_{ijk}\hat{x}_j^L \hat{P}_k\ , \nonumber \\
& &\hat{P}_1=-\frac{1}{\theta}\ad{\hat{x}^0}\ ,\ \hat{P}_a =
-i\frac{\partial}{\partial \hat{x}_a}\ \ (a=2,3)\ , \nonumber \\
& &[\hat{L}_i,\hat{L}_j]=i\varepsilon_{ijk}\hat{L}_k \ .
\end{eqnarray}
The coordinates $\hat{x}_i^L$ and momenta $\hat{P}_i$ respond to
$\hat{L}_i$ as they should to infinitesimal rotations:
\begin{eqnarray}
& &[\hat{L}_i , \hat{x}^L_j] = i \varepsilon_{ijk}\hat{x}^L_k\ , \nonumber \\
& &[\hat{L}_i , \hat{P}_j] = i \varepsilon_{ijk}\hat{P}_k\ .
\nonumber
\end{eqnarray}
But still, we cannot regard $\hat{L}_i$ as generating spatial
rotations as it affects $\hat{x}_0^L$ as well: \be
\label{Time}[\hat{L}_i , \hat{x}^L_0] = i
\theta\varepsilon_{1ik}\hat{P}_k\ . \ee

We should expect this result as the algebra
$\mathcal{A}_\theta(\mathbb{R}^4)$ does not admit spatial
rotations as automorphisms:
$$
[\hat{x}_0,R_{1j}\hat{x}_j]\ne i\theta \mbox{ for all } R\in
SO(3)\ .
$$

Now suppose that the Hamiltonian $H$ for $\theta_{0i}=0$ is
time-independent and invariant under rotations. It may have
eigenstates $\varphi_E^{(n)}$, $n\in [1,...,N]$ degenerate in
energy and carrying a representation of the symmetry group
$SO(3)$. Then by the spectral theorem, $\hat{H}$ for
$\theta_{0i}\ne 0$ also has this energy degeneracy and eigenstates
\begin{eqnarray}
& &\hat\psi_E
^{(n)}={\varphi}_E^{(n)}(\vec{\hat{x}})\mathrm{e}^{(-iE
\hat{x}_0)}\ , \nonumber \\
& &\hat{H}\hat\psi_E^{(n)} = E\hat\psi_E^{(n)}\ . \nonumber
\end{eqnarray}
Here we have represented ${\varphi}_E^{(n)}$ as a function of
spatial coordinates.

The mechanical reason for the persistence of degeneracies for
$\theta_{0i}\ne 0$ is thus clear. But can we locate an underlying
noncommutative symmetry?

We consider $H$ invariant under rotations:
$$
[L_i , H](\vec{x},\vec{p})=0\ .
$$
Then
\begin{eqnarray}
& &[\hat{L}_i , \hat{H}]=[L_i , H](\vec{x},\vec{p})|_{x_i=\hat{x}_i^L,p_i=\hat{P}_i}=0\ , \nonumber \\
& &[\hat{L}_i , \hat{P}_0]=0\ . \nonumber
\end{eqnarray}

Thus the group $SO(3)$ generated by $\hat{L}_i$ preserves
$\hat{H}$ and the Schr\"{o}dinger constraint: it is a
noncommutative symmetry group. Furthermore its action on energy
eigenstates is something familiar:
$$
\hat{L}_i\hat\psi_E
^{(n)}=\left(L_i{\varphi}_E^{(n)}\right)(\vec{\hat{x}})|_{x_i=\hat{x}_i}\mathrm{e}^{(-iE
\hat{x}_0)}\ .
$$

In this way we see that the noncommutative $SO(3)$ can explain
spectral degeneracies even though this $SO(3)$ is not the spatial
rotation group.

The noncommutative $SO(3)$ is \textit{not} a symmetry if the
Hamiltonian $H$ for $\theta_{0i}=0$, although commuting with
$L_i$, has explicit time-dependence:
\begin{eqnarray}
& &H=H(x_0,\vec{x},\vec{p})\ , \nonumber \\
& &[{L}_i , H]=0\ . \nonumber
\end{eqnarray}
In that case
$$
\hat{H}=\hat{H}(\hat{x}_0^L,\vec{\hat{x}}^L,\vec{\hat{P}})
$$
and
$$
[\hat{L}_i , \hat{H}]\ne 0
$$
because of Eq.(\ref{Time}).

Thus effects of noncommutativity on spatial rotations are revealed
only by time-dependent $\hat{H}$.

\noindent\textit{iii) Noncommutative Parity and its Action as a
Symmetry}

If $H$ is a time-independent Hamiltonian for $\theta_{0i}=0$ which
is invariant under parity $P$,
$$
PH(\vec{x},\vec{p})P^{-1}=H(-\vec{x},-\vec{p})=H(\vec{x},\vec{p})\
,
$$
it so happens that there is a deformed noncommutative parity
$\hat{P}$ which is a noncommutative symmetry. But it affects time
$\hat{x}_0^L$ and is not properly spatial reflection. Still it is
a valid symmetry and good for explaining energy degeneracies.

$\hat{P}$ can be constructed as follows. Let $\mathcal{P}_\theta$
be the plane perpendicular to $\vec\theta_0$:
$$
\vec{x}\in \mathcal{P}_\theta \ \Leftrightarrow \ \vec{x}\cdot
\vec\theta_0 = 0\ .
$$
It is spanned by an orthonormal basis $\vec{e}^{(a)},\ a=(1,2)$:
$$
\vec{e}^{(a)}\cdot\vec{e}^{(b)}=\delta_{ab}\ , \
\vec\theta_0\cdot\vec{e}^{(a)}=0\ .
$$
We can write
\begin{eqnarray}
& &\vec{x}=\vec{e}^{(a)}(\vec{x}\cdot\vec{e}^{(a)}) +
\frac{\vec\theta_0}{|\vec\theta_0|}(\vec\theta_0\cdot\vec{x})\ , \nonumber \\
& &|\vec\theta_0|=\left|\left(\sum_i
\theta_{0i}^2\right)^{1/2}\right|\ . \nonumber
\end{eqnarray}

Let $K$ be the operator of reflection of just
$\vec{x}\cdot\vec{e}^{(1)}$ in the commutative case:
\begin{eqnarray}
& &K\vec{x}\cdot\vec{e}^{(1)}K^{-1}=-\vec{x}\cdot\vec{e}^{(1)}\ , \nonumber \\
& &K\vec{x}\cdot\vec{e}^{(2)}K^{-1}=\vec{x}\cdot\vec{e}^{(2)}\ ,\
K\vec{x}\cdot\vec\theta_0K^{-1}=\vec{x}\cdot\vec\theta_0\ .
\nonumber
\end{eqnarray}
Then commutative parity $P$ is $R_{\vec{e}^{(1)}}(\pi)K$ where
$R_{\vec{e}^{(1)}}(\pi)$ is rotation by $\pi$ around
$\vec{e}^{(1)}$-axis.

As remarked earlier, $\hat{K}$, the noncommutative version of $K$,
is an automorphism of $\mathcal{A}_\theta(\mathbb{R}^4)$. The
noncommutative version $\hat{R}_{\vec{e}^{(1)}}(\pi)$ of
$R_{\vec{e}^{(1)}}(\pi)$ is well-defined as well: it is an element
of the $SO(3)$ group with generators $\hat{L}_i$. The
noncommutative parity is thus
$$
\hat{P}=\hat{R}_{\vec{e}^{(1)}}(\pi)\hat{K}\ .
$$
We have
$$
\hat{P}\hat{x}^L_i\hat{P}^{-1}=-\hat{x}^L_i\ ,\
\hat{P}\hat{P}_i\hat{P}^{-1}=-\hat{P}_i\ .
$$

$\hat{P}$ affects $\hat{x}_0^L$ because
$\hat{R}_{\vec{e}^{(1)}}(\pi)$ does so. Hence it cannot be
regarded as just total spatial reflection. However it does not
affect $\hat{x}_0^R$.

$\hat{P}$ does not depend on the choice of the axis
$\vec{e}^{(1)}$ in the plane perpendicular to $\vec\theta_0$. For
example if
$$
\hat{P}'=\hat{R}_{\vec{e}^{(2)}}(\pi)\hat{K}'\ ,
$$
where $\hat{K}'$ reflects $\vec{x}\cdot\vec{e}^{(2)}$, then
\be\label{Parities} \hat{P}'=\hat{P}\ .\ee

The proof of Eq.(\ref{Parities}) is as follows. Let
$$
U = \hat{P}'^{-1}\hat{P}\ .
$$
Then
$$
U\hat{x}^L_i U^{-1}=\hat{x}^L_i\ ,\ U\hat{P}_i U^{-1}=\hat{P}_i\
,\ U\hat{x}^R_i U^{-1}=\hat{x}^R_i\ .
$$
Hence
$$
U\hat{x}^L_0 U^{-1}=U[-\theta \hat{P}_1 +\hat{x}^R_0]
U^{-1}=\hat{x}^L_0\ .
$$
Thus since conjugation by $U$ affects no operator, we can identify
$\hat{P}'$ with $\hat{P}$.

Since $\hat{R}_{\vec{e}^{(1)}}(\pi)$ and $\hat{K}$ commute with
$\hat{P}_0$, so does $\hat{P}$. It follows as before that if $P$
commutes with $H$ and $H$ is time-independent, $\hat{P}$ commutes
with $\hat{H}$ and also preserves the Schr\"{o}dinger constraint.

Thus degeneracies due to parity in commutative quantum physics are
preserved intact in noncommutative quantum physics if $H$ is
time-independent.

But $\hat{P}$ is not spatial reflection as it affects coordinate
time: $\hat{P}\hat{x}_0^L\hat{P}^{-1}\ne \hat{x}_0^L$. It is
rather a `noncommutative' or `deformed' parity.

\section{Forbidden Transition in Hydrogen Atom: A Quantum Field Theory Example.}

We saw that energy levels of time-independent Hamiltonians in
quantum mechanics cannot reveal effects of noncommutativity. We
can examine time-dependent processes such as transitions between
levels to see the effects of the latter. Other alternatives are
interference phenomena \cite{BKS}. We focus on the former here.

As an example of the transition induced by the noncommutative
$\theta$-parameter, we examine the one-photon transition
$2s\rightarrow 1s +\gamma$ in hydrogen. (For discussion of the
effects of space-space noncommutativity in hydrogen atom, see
\cite{Chaichian:2000si}.) It is forbidden for $\theta = 0$ in the
absence of electron spin effects, but can occur for $\theta \ne
0$. It would be thus a genuine $\theta$-effect. But the amplitude
vanishes at tree level for the same reason (standard rotational
invariance) that it vanishes for $\theta = 0$. The one-loop
amplitude is sensitive to the breakdown of standard rotational
symmetry and is not zero. However it is extremely small even when
$\theta$ is of the order of $(TeV)^{-2}$. This transition in
hydrogen is not a realistic process to detect or bound $\theta$,
as backgrounds, including the magnetic $2s\rightarrow 1s + \gamma$
transition, will overwhelm effects of $\theta$. So we do not
present the one-loop calculation here.

In the final section, we speculate on more realistic processes to
detect $\theta$.

\noindent\textit{i) On Relative Coordinates.}\footnote{This
section is based on work by A.P.B. with Sachin Vaidya.}

We have to clarify a conceptual issue before beginning the study
of the process $2s\rightarrow 1s + \gamma$.

In conventional physics, when $\theta = 0$ and there are $N$
non-identical particles moving in $\mathbb{R}^3$, the
configuration space is $\mathbb{R}^{3N}$
\cite{Balachandran:1991zj}. The corresponding coordinate function
$\hat{x}^{(m)}$ ($m=1,...,N$) of the $m^{th}$ particle is defined
by
$$
\hat{x}^{(m)}_j(\vec{x}^{(1)},\vec{x}^{(2)},...,\vec{x}^{(N)}) =
x^{(m)}_j\ .
$$
There is only \textit{one} time operator $\hat{x}^{0}$ common to
all particles.

The $\theta \ne 0$ generalization of this space-time algebra has
the same commutator of $\hat{x}_{0}$ with \textit{all}
$\vec{\hat{x}}^{(m)}$ (see Eq.(\ref{CR0})),
$$
[\hat{x}_{0},\hat{x}^{(m)}_{j}] = i\theta s_i
$$
while spatial coordinate functions have vanishing commutators:
$$
[\hat{x}^{(m)}_{i},\hat{x}^{(n)}_{j}] = 0\ ,\ \ m,n\in
[1,2,...,N],\ i,j\in [1,2,3]\ .
$$

It follows that \textit{relative} spatial coordinates
\textit{commute} with $\hat{x}^{0}$:
$$
[\hat{x}_{0},\hat{x}^{(m)}_{i} - \hat{x}^{(n)}_{i}]=0\ .
$$
The algebra generated by $\hat{x}_{0}$, $\hat{x}^{(m)}_{i} -
\hat{x}^{(n)}_{i}$ is not sensitive to $\theta$. It is just a
commutative algebra.

We can describe the situation in another way. Let us associate the
$\theta\ne 0$ space-time algebra of $\hat{x}_{0}$, $\vec{\hat{x}}$
to one of the particles, say 1, chosen at random. Then the
space-time algebra of $m^{th}$ particle is obtained by spatial
translations: It has generators $\hat{x}_{0}$, $\hat{x}_{i} +
\hat{a}^{(m)}_{i}$, where $\hat{a}^{(m)}_{i}$ is the relative
coordinate $\hat{x}^{(m)}_{i} - \hat{x}^{(1)}_{i}$.

For this reason, spatial rotations act in a standard way, with
angular momenta $L_i = -i(\vec{\hat{a}}\wedge\vec\nabla )_i$,
$\nabla_i = \frac{\partial}{\partial\hat{a}_{i}}$, on relative
coordinates. Now suppose the commutative Hamiltonian has
rotational symmetry. Then its noncommutative version restricted to
relative coordinate-time algebra would also have that symmetry,
unless technical problems like factor-ordering interfere. We need
a nontrivial presence of the noncommutative algebra of the center
of mass before $\theta$-effects show up.

For such reasons, the $\theta$-effect does not show up in the
process $2s \rightarrow 1s +\gamma $ at tree level. But it does
show up at one loop.

\noindent\textit{ii) The Fields and the Hamiltonian.}

Quantum field theory (QFT) gives a conceptually clean approach to
study our process.

For $\theta\ne 0$, the second quantized (free) photon field has
the mode expansion
\begin{eqnarray} \label{A}
& &{\hat{{A}}}_i (\vec{\hat{x}},\hat{x}_0) =\int
\frac{d^3k}{(2\pi)^{3/2}\sqrt{2\omega}}\sum_\alpha \left(
\mathrm{a}(\vec{k},\alpha){\epsilon}_i^{(\alpha )}(\vec{k})
\mathrm{e}^{i\vec{k}\cdot\vec{\hat{x}}}
\mathrm{e}^{-i\omega\hat{x}_0} +
\mathrm{e}^{i\omega\hat{x}_0}\mathrm{e}^{-i\vec{k}\cdot\vec{\hat{x}}}
{\bar\epsilon}_i^{(\alpha
)}(\vec{k})\mathrm{a}(\vec{k},\alpha)^\dagger \right)\
,\nonumber\\
&
&[\mathrm{a}(\vec{k},\alpha),\mathrm{a}(\vec{k}',\beta)^\dagger]=\delta^3
(k'-k)\delta_{\alpha\beta}\ \mathrm{etc.},\\
& & \omega = |\vec{k}|\ ,\nonumber
\end{eqnarray}
$\alpha$ denoting photon polarization. We work in radiation gauge:
\be \label{gauge}
\vec{k}\cdot\vec{\epsilon}^{(\alpha)}(\vec{k})=0\ . \ee

As for the hydrogen atom, we denote the electron-proton relative
coordinate by $\vec{a}$. As we discussed, it commutes with
$\hat{x}^0$. If the electron and proton have masses $m$ and $M$
and coordinates $\vec{\hat{x}}^{(e)}$ and $\vec{\hat{x}}^{(p)}$,
the center of mass coordinate is
$$
\vec{\hat{x}}=\frac{m\vec{\hat{x}}^{(e)}+M\vec{\hat{x}}^{(p)}}{m+M}\
.
$$
It has the commutator
$$
[\hat{x}_0 , \hat{x}_i]=i\theta \delta_{1i}
$$
with $\hat{x}_0$.

The hydrogen atom bound state wave functions for energies $E_n$
can be denoted by $\phi_n$ and continuum wave functions of energy
$E$ by $\phi_E$. (We ignore spin effects. $n$ is a discrete level
and $E$ is energy.) They are functions of $\vec{a}$ and have the
normalization
$$
(\phi_n , \phi_m)=\int d^3a\ \bar\phi_n (a) \phi_m (a) =
\delta_{nm}\ ,
$$
$$
(\phi_E , \phi_{E'})=\delta (E-E')\ , \mathrm{etc.}
$$

The second-quantized (non-relativistic) hydrogen field $\hat\Psi$
is given by
$$
\hat\Psi (\vec{\hat{x}},\vec{a},\hat{x}_0)=\int\frac{d^3
p}{(2\pi)^{3/2}}\left[ \sum_n\mathrm{a}_n (\vec{p})\phi_n
(\vec{a})\mathrm{e}^{-iE_n\hat{x}_0}+\int dE\ \mathrm{a}_E
(\vec{p})\phi_E (\vec{a})\mathrm{e}^{-iE\hat{x}_0}
\right]\mathrm{e}^{i\vec{p}\cdot\vec{\hat{x}}}\mathrm{e}^{-i\frac{{p}^2}{2(m+M)}
\hat{x}_0}\ ,
$$
$$
[\mathrm{a}_n (\vec{p}),\mathrm{a}_m (\vec{p}\
')^\dagger]=\delta_{nm}\delta^3 (\vec{p}-\vec{p}\ ')\ ,
$$
$$
[\mathrm{a}_E (\vec{p}),\mathrm{a}_{E'} (\vec{p}\
')^\dagger]=\delta (E-E')\delta^3 (\vec{p}-\vec{p}\ ')\
\mathrm{etc.}
$$
The labels $\vec{p}$, $\vec{p}'$ and the factor
$\mathrm{e}^{i\vec{p}\cdot\vec{\hat{x}}}\mathrm{e}^{-i\frac{{p}^2}{2(m+M)}
\hat{x}_0}$ come from center-of-mass motion.

For purposes of illustration, it is enough to couple the photon
field just to the electron. The single-particle interaction
Hamiltonian linear in $\vec{\hat{A}}$ is then
$$
-e\left(\vec{\hat{P}}\cdot \vec{\hat{A}}(\hat{x}^{(e)}) +
\vec{\hat{A}}(\hat{x}^{(e)})\cdot
\vec{\hat{P}}\right)=-2e\vec{\hat{A}}(\hat{x}^{(e)})\cdot
\vec{\hat{P}} \ ,\ \hat{P}_i = -i\frac{\partial}{\partial
x^{(e)}_i}
$$
in view of (\ref{gauge}). Here $\vec{\hat{x}}^{(e)}=\vec{\hat{x}}
+ (1-\mu)\vec{\hat{a}}$ and $\mu=\frac{m}{m+M}$ is the reduced
mass. The QFT interaction Hamiltonian is thus
$$
H_I = -2e\int d^3 a\ S\left(\hat{\Psi}^\dagger
\vec{\hat{A}}(\vec{\hat{x}} + (1-\mu)\vec{\hat{a}})\cdot
\vec{\hat{P}}\hat{\Psi}\right)\ ,
$$
where the positive map $S$ refers to the algebra of $\hat{x}_\mu$
and it is to be evaluated at some time $x_0$ (the value of $x_0$
does not affect final answers).

The free Hamiltonian $H_0$ is the sum of those for hydrogen and
photon. The interaction representation $S$-matrix is
$$
\mathcal{S} = T \exp \left( -i\int^{\infty}_{-\infty}d\tau\
U_0^{-1}(\tau)H_I U_0 (\tau)\right)\ ,
$$
$$
U_0 (\tau)=\exp (-i\tau H_0)\ .
$$

When the $2s$ level at rest decays into $1s + photon\ \gamma $,
the momenta of $1s$ and $\gamma$ being $\pm \vec{k}$ and photon
helicity being $\lambda$, the first order transition matrix
element is \be \label{t1}T^{(1)} = -i\int^{\infty}_{-\infty}d\tau\
\langle 1s(\vec{k}),\lambda (-\vec{k})|e^{i\tau H_0}H_I e^{-i\tau
H_0}|2s \rangle \ . \ee

We now isolate the integral involving relative coordinate here and
show that it vanishes.

As the final state involves photon of momentum $-\vec{k}$ and
helicity $\lambda$, the component \be \label{comp}
\mathrm{a}(-\vec{k},\lambda){\epsilon}^{(\lambda
)}_i(-\vec{k})e^{-i\omega\hat{x}_0}e^{-i\vec{k}\cdot
\vec{\hat{x}}^{(e)}}=\mathrm{a}(-\vec{k},\lambda){\epsilon}^{(\lambda
)}_i(-\vec{k})e^{-i\omega\hat{x}_0}e^{-i\vec{k}\cdot
[(\vec{\hat{x}} + (1-\mu)\vec{\hat{a}}]} \ee of $\hat{A}_i$ is
picked out in the matrix element of (\ref{t1}). It gets multiplied
by $\hat{P}_i\Psi$. But the $1s$ state has momentum $\vec{k}$ so
that $\hat{P}_i\Psi$ has factor $k_i$. Since
$\vec{\epsilon}^{\lambda}(-\vec{k})\cdot \vec{k} = 0$ the entire
matrix element vanishes:
$$
T^{(1)}=0\ .
$$

As the $S$-matrix has been presented in the second-quantized
formalism, the process $2s\rightarrow 1s + \gamma$ can be
investigated beyond the tree approximation.

\section{Discussion: The Decay $Z^0\rightarrow 2\gamma$}

As we saw in previous sections, selection rules from rotational
symmetry for $\theta_{0i}=0$ are not in general respected in
scattering and decay processes when $\theta_{0i}\ne 0$. One
candidate for such a process is the decay of a massive vector
particle into two photons, such as $Z^0\rightarrow 2\gamma$.
Though one can easily write an effective Lagrangian density
$L_{int}$ for this process, the resulting amplitude is zero if
$\theta_{0i}=0$. For example, let us consider $L_{int}\sim
{F}_{\mu\nu}G^{\nu\rho}(*F)_\rho^{\,\mu}$, where
${F}_{\mu\nu}=\partial_\mu A_\nu-\partial_\nu A_\mu$,
${G}_{\mu\nu}=\partial_\mu B_\nu-\partial_\nu B_\mu$ and
$(*F)_{\mu\nu}$ is dual of ${F}_{\mu\nu}$. ${A}_\mu$ and ${B}_\mu$
are massless and massive vector fields
respectively.\footnote{Another choice is $L_{int}\sim
F_{\mu\nu}G^{\nu\rho}F_\rho^{\,\mu}$. But this is zero just due to
the antisymmetry of ${F}_{\mu\nu}$ and ${G}_{\mu\nu}$.} The decay
amplitude is then proportional to \be \label{Ampl}\mathcal{A}\sim
\varepsilon^{\rho\mu\eta\gamma}k'_\eta \epsilon'_\gamma
(k_\mu\epsilon^\nu - k^\nu\epsilon_\mu)(p_\nu \varepsilon_\rho -
p_\rho \varepsilon_\nu)\ , \ee where $k_\mu,\ \epsilon_\mu$ and
$k'_\mu,\ \epsilon'_\mu$ are momentum and polarization of photons
and $p_\mu,\ \varepsilon_\mu$ those of the massive vector
particle. Calculation shows that this is zero upon using the
transversality conditions on the polarization vectors. This result
holds in general.

The consideration of the above process with time-space
noncommutativity requires a better understanding of quantum theory
when $\theta_{0i}\ne 0$. The work under progress indicates that it
would occur when $\theta_{0i}\ne 0$.

In conclusion, in this paper we have considered the effects of
time-space noncommutativity due to deformation of the rotation
symmetry and parity in the case of nonzero $\theta_{0i}$. It is
argued that many processes that are forbidden in the commutative
case become allowed by this deformation. Our point is supported by
the explicit calculation of the decay rate of the transition
$2s\rightarrow 1s + 2\gamma$ in hydrogen atom. Comments on the
processes like $Z^0\rightarrow 2\gamma$ have also been made.

\bigskip
\bigskip

{\parindent 0cm{\bf Acknowledgement}}

We are very grateful to J. Schechter, S. Vaidya, J.K.
Bhattacharjee, A. Honig and S. Kurkcuoglu for discussions. We
especially thank H.~S.~Mani for drawing A.~P.~B.'s attention to
the $Z^0\rightarrow 2\gamma$ process and suggesting its
significance for our considerations. This work was supported in
part by the DOE grant DE-FG02-85ER40231 and by NSF under contract
number INT9908763.


\bigskip
\bigskip


\begin{thebibliography}{99}

%\cite{Doplicher:1994tu}
\bibitem{Doplicher:1994tu}
S.~Doplicher, K.~Fredenhagen and J.~E.~Roberts,
%``The Quantum structure of space-time at the Planck scale and quantum fields,''
Commun.\ Math.\ Phys.\  {\bf 172}, 187 (1995)
[arXiv:hep-th/0303037].
%%CITATION = HEP-TH 0303037;%%

%\cite{Gomis:2000zz}
%\bibitem{Gomis:2000zz} J.~Gomis and T.~Mehen,
%``Space-time noncommutative field theories and unitarity,''
%Nucl.\ Phys.\ B {\bf 591}, 265 (2000) [arXiv:hep-th/0005129].
%%CITATION = HEP-TH 0005129;%%

%\cite{Seiberg:1999vs}
\bibitem{Seiberg:1999vs}
N.~Seiberg and E.~Witten,
%``String theory and noncommutative geometry,''
JHEP {\bf 9909}, 032 (1999) [arXiv:hep-th/9908142].
%%CITATION = HEP-TH 9908142;%%

%\cite{Yang:1950rg}
\bibitem{Yang:1950rg}
C.~N.~Yang,
%``Selection Rules For The Dematerialization Of A Particle Into Two Photons,''
Phys.\ Rev.\  {\bf 77}, 242 (1950).
%%CITATION = PHRVA,77,242;%%

%\cite{Behr:2002wx}
\bibitem{Behr:2002wx}
W.~Behr, N.~G.~Deshpande, G.~Duplancic, P.~Schupp, J.~Trampetic
and J.~Wess,
%``The Z $\to$ gamma gamma, g g decays in the noncommutative standard model,''
Eur.\ Phys.\ J.\ C {\bf 29}, 441 (2003) [arXiv:hep-ph/0202121].
%%CITATION = HEP-PH 0202121;%%

%\cite{Bahns:2002vm}
%\bibitem{Bahns:2002vm}
%D.~Bahns, S.~Doplicher, K.~Fredenhagen and G.~Piacitelli,
%``On the unitarity problem in space/time noncommutative theories,''
%Phys.\ Lett.\ B {\bf 533} (2002) 178 [arXiv:hep-th/0201222].
%%CITATION = HEP-TH 0201222;%%

%\cite{Balachandran:2004rq}
\bibitem{Balachandran:2004rq}
A.~P.~Balachandran, T.~R.~Govindarajan, C.~Molina and
P.~Teotonio-Sobrinho,
%``Unitary quantum physics with time-space noncommutativity,''
arXiv:hep-th/0406125.
%%CITATION = HEP-TH 0406125;%%



%\cite{Pinzul:2004qu}
%\bibitem{Pinzul:2004qu}
%A.~Pinzul and A.~Stern,
%``Space-time noncommutativity from particle mechanics,''
%arXiv:hep-th/0402220.
%%CITATION = HEP-TH 0402220;%%

%\cite{Banerjee:2004ms}
%\bibitem{Banerjee:2004ms}
%R.~Banerjee, B.~Chakraborty and S.~Gangopadhyay,
%``Reparametrisation symmetry and Noncommutativity in particle mechanics,''
%arXiv:hep-th/0405178.
%%CITATION = HEP-TH 0405178;%%

%\cite{Hinchliffe:2002km}
\bibitem{Hinchliffe:2002km}
I.~Hinchliffe and N.~Kersting,
%``Review of the phenomenology of noncommutative geometry,''
Int.\ J.\ Mod.\ Phys.\ A {\bf 19}, 179 (2004)
[arXiv:hep-ph/0205040].
%%CITATION = HEP-PH 0205040;%%
%\cite{Chaichian:2004za}

\bibitem{Chaichian:2004za}
M.~Chaichian, P.~P.~Kulish, K.~Nishijima and A.~Tureanu,
%``On a Lorentz-invariant interpretation of noncommutative space-time and its
%implications on noncommutative QFT,''
arXiv:hep-th/0408069;
%%CITATION = HEP-TH 0408069;%%
%\cite{Chaichian:2004yh}
%\bibitem{Chaichian:2004yh}
M.~Chaichian, P.~Presnajder and A.~Tureanu,
%``New concept of relativistic invariance in NC space-time: Twisted Poincare
%symmetry and its implications,''
arXiv:hep-th/0409096.

%%CITATION = HEP-TH 0409096;%%
%\cite{Alexanian:2000uz}
%\bibitem{Alexanian:2000uz}
%G.~Alexanian, A.~Pinzul and A.~Stern,
%``Generalized Coherent State Approach to Star Products and Applications to the
%Fuzzy Sphere,''
%Nucl.\ Phys.\ B {\bf 600}, 531 (2001) [arXiv:hep-th/0010187].
%%CITATION = HEP-TH 0010187;%%

\bibitem{BKS}
A.~P.~Balachandran, K.~S.~Gupta and S.~Kurkcuoglu, work in
progress.

%\cite{Chaichian:2000si}
\bibitem{Chaichian:2000si}
M.~Chaichian, M.~M.~Sheikh-Jabbari and A.~Tureanu,
%``Hydrogen atom spectrum and the Lamb shift in noncommutative QED,''
Phys.\ Rev.\ Lett.\  {\bf 86}, 2716 (2001) [arXiv:hep-th/0010175].
%%CITATION = HEP-TH 0010175;%%

%\cite{Balachandran:1991zj}
\bibitem{Balachandran:1991zj}
A.~P.~Balachandran, G.~Marmo, B.~S.~Skagerstam and A.~Stern,
``Classical topology and quantum states'', Singapore, World
Scientific (1991) 358 p.





\end{thebibliography}
\end{document}